\documentclass{elsart}
\input psfig.tex

\def\ltsima{$\; \buildrel < \over \sim \;$}
\def\simlt{\lower.5ex\hbox{\ltsima}}		
\def\gtsima{$\; \buildrel > \over \sim \;$}
\def\simgt{\lower.5ex\hbox{\gtsima}}		

\begin{document}
\runauthor{Urry}
\begin{frontmatter}

\title{ Multiwavelength Properties of Blazars}
\author[STScI]{C. Megan Urry}
\address[STScI]{Space Telescope Science Institute, Baltimore, Maryland}
\begin{abstract}
Blazar spectral energy distributions (SEDs) are double peaked and
follow a self-similar sequence in luminosity. The so-called ``blue"
blazars, whose first SED component peaks at X-ray energies, are
TeV sources, although with a relatively small fraction of their
bolometric luminosities. The ``red" blazars, with SED peaks
in the infrared-optical range, appear to emit relatively more power
in the gamma-ray component but at much lower energies (GeV and below).
Correlated variations across the SEDs (of both types) are consistent
with the picture that a single electron population
gives rise to the high-energy parts of both SED components, 
via synchrotron at low
energies and Compton-scattering at high energies. In this
scenario, the trends of SED shape with luminosity can be explained
by electron cooling on ambient photons. With simple assumptions,
we can make some estimates of the physical conditions in blazar
jets of each ``type" and can predict which blazars are the most
likely TeV sources. Upper limits from a mini-survey of candidate TeV
sources indicate that only $\sim10$\% of their bolometric luminosity is
radiated in gamma-rays, assuming the two SED components peak near
1~keV and 1~TeV. 
Finally, present blazar samples are too shallow to indicate what kinds
of jets nature prefers, i.e., whether the low-luminosity ``blue"
blazars or the high-luminosity ``red" blazars are more common.
\end{abstract}
\begin{keyword}
blazars; BL Lac objects; multiwavelength spectra
\end{keyword}
\end{frontmatter}

\section{The Blazar Paradigm}

\subsection{SEDs and variability}

The spectral energy distributions (SEDs) of blazars have two components,
with peaks (in the usual $\nu F_\nu$ representation) separated by
$\sim8-10$ decades in frequency, as discussed by \cite{UMU,Foss} and 
several authors in this volume. 
The lower frequency component has peak frequency, $\nu_{p}$,
in the IR-X-ray energy range, and is generally believed to be due to 
synchrotron radiation, so the peak can be identified with a characteristic
electron energy $\gamma_{peak} \propto \sqrt{\nu_{peak}/B}$.
Typically blazars\footnote{We use the term blazar as the collective
name for BL Lac objects and flat-spectrum radio-loud quasars (also
known in most cases as HPQs, or highly polarized quasars). All are
highly variable, polarized, and have strong radio emission from
compact flat-spectrum cores.}
are more variable --- more rapidly and with larger
amplitude --- at or above the spectral peak, certainly in the synchrotron
component and plausibly in the gamma-ray component, although here the
data are much more sparse. 

The variability in the two blazar spectral components is correlated, 
with wavelength-dependent lags,
between approximately corresponding points on the two curves 
(\cite{UMU}).
This has led to the suggestion that a single population of electrons could
be responsible for the bulk of the emission in both components, 
via synchrotron radiation in the low energy bump and via Compton 
scattering of local soft photons in the high energy bump.
(Other components contribute to the extended radio emission and
to an optical/UV bump, if present.)
For the remainder of the talk, I refer to ``synchrotron'' and 
``Compton'' spectral components, but recognize that for the 
high-energy emission at least, there are other still-plausible
models.

As others discuss at this conference, the origin of the seed photons 
for the Compton-scattered gamma-rays has been a subject of some
speculation. Different possibilities include local synchrotron 
photons produced by the same electron population (the SSC model,
\cite{M92}),
or sources external to the jet, such as accretion disk photons
(\cite{DS}) or reprocessed photons from the
broad-line region (\cite{SBR}). 
Quite possibly the relative importance of these possibilities
changes from one source to the next, or between active and quiescent
periods in a given source, as the intrinsic source
properties change (\cite{GM}).

There are clues in the trends in blazar properties with
luminosity. Rita Sambruna showed in her Ph.D. thesis 
that synchrotron peak frequencies and overall spectral shapes differed 
among X-ray-selected BL Lac objects, radio-selected BL Lac objects,
and flat-spectrum radio quasars (\cite{SMU}),
in a way that could not be explained simply by beaming. 
Instead, the SEDs changed shape systematically with luminosity,
such that high luminosity BL Lac objects and quasars have synchrotron
peaks in the IR-optical, while lower luminosity BL Lacs peak at
higher energies, typically UV-X-ray. 
We refer to these as LBL (for low-frequency peaked blazars) and 
HBL (for high-frequency peaked blazars), respectively (see \cite{PG95}
for exact definition), or more descriptively, as ``red'' and ``blue'' blazars. 
BL Lacs found in the first X-ray surveys were generally HBL/blue;
BL Lacs found in classical radio surveys were generally LBL/red, as
were most FSRQ (flat-spectrum radio-loud quasars), though ``bluer'' FSRQ 
are now being identified in multiwavelength surveys designed not to exclude
them (\cite{P98,Sally}). 

\begin{center}
\begin{table}
\caption{Spectral Trends in Blazars} 
\label{tab:trends} 
\begin{tabular}{cc}
\hline 
	as $L_{bolometric}$ increases, &$\nu_{synchrotron}$ decreases \\
				&$L_{Compton}$ increases \\
				&$L_{\rm emission-line}$ increases \\
\multicolumn{2}{c}{$\nu_{Compton}/\nu_{synchrotron}$ 
roughly similar in HBL, LBL, FSRQ}\\
\hline 
\end{tabular}
\end{table}
\end{center}

Taking a synthetic view of the origin of the SEDs, it seems
the bright EGRET blazars, which have peak gamma-ray emission 
near $\sim1$~GeV, are much like the few known TeV blazars, 
which peak $\simgt100$ times higher in energy. This is important
because many more ``GeV'' blazars are known and have been studied 
than ``TeV'' blazars. With appropriate scaling, we should be able
to use the observed behavior of the GeV blazars (see \cite{aew2})
to inform our multiwavelength studies of TeV blazars 
(discussed below and see \cite{tad2}).

\subsection{Compton cooling}

The SED trends support a simple paradigm 
(\cite{Foss,G98}) in which electrons in higher luminosity blazars 
suffer more cooling because of larger external photon densities in
the jet (as indicated by the higher $L_{\rm emission-line}$ characteristic 
of this group).
This leads naturally to lower characteristic electron energies,
as well as larger ratios of inverse Compton to synchrotron luminosity
($L_C/L_s$) because of the ``external
Compton'' (EC) contribution. In contrast, lower luminosity blazars are
dominated by (weaker) SSC cooling, and so have electrons with 
higher $\gamma_{peak}$. 
The correlation of $\gamma_{peak}$ with energy density may even suggest 
that electron acceleration in blazars is independent of luminosity,
and that only the cooling differs (\cite{G98,ghis2}).
A similar luminosity-linked scheme, with a somewhat more physical
motivation, was developed by Markos Georganopoulos in his Ph.D. thesis
(\cite{GeorgB}).
In any case, the current view
is that the EC process dominates the gamma-ray production in high-luminosity 
blazars, while the SSC process dominates in low-luminosity blazars.
The X-rays from either high- or low-luminosity blazars are
likely dominated by SSC (\cite{Kubo,tad2}).

The similarity of $\nu_C/\nu_s$ in HBL and FSRQ/LBL is somewhat puzzling 
in this picture, and will be physically important if it persists when (if?)
further TeV observations define the HBL Compton peaks well (this is
complicated by the effect of the Klein-Nishina cross section at TeV energies).
In a simple homogeneous SSC model, $\nu_C/\nu_s \propto \gamma_{peak}^2$, 
whereas in the simplest EC models $\nu_C/\nu_s \propto \nu_E/B$,
where $\nu_E$ is the characteristic frequency of the external seed photons.
If $\nu_C/\nu_s$ changes little from SSC-dominated to EC-dominated
blazars, then perhaps energy is distributed so that the typical electron
energy, which depends on acceleration, cooling, and escape, maintains
the appropriate relation to the magnetic field energy density.

\subsection{Caveats}

Two caveats to this scheme bear mentioning, as
both can lead to higher {\it observed} $L_C/L_s$ in high-luminosity blazars,
even when there are no significant intrinsic differences.
First, Dermer has pointed out that EC gamma-rays are more tightly beamed than
SSC. Therefore, if the EC process dominates the gamma-ray emission
in high-luminosity blazars but the SSC process dominates in low-luminosity
blazars, the increase of $L_C/L_s$ with luminosity will be exaggerated
(and could even be spurious).

Second is a sort of ``variability bias'' (see \cite{Wagner,Hart2}).
Given the limited sensitivity of EGRET, which
typically had to integrate for one or more weeks to detect a blazar, 
this bias arises if two plausible assumptions hold:
(1) high-luminosity sources are larger than low-luminosity sources, 
and their intrinsic variability time scales are proportionately longer, 
and (2) blazar gamma-ray light curves are characterized by flares 
separated by quiescent periods, as suggested by the long-term EGRET
light curve of 3C~279 (\cite{Hart1}).
Scaling from 3C~279, even as slowly as $L^{-1/2}$,
a typical EGRET observation would span several 
flares in a low-luminosity blazar, sampling an average flux close to 
the quiescent value, hence implying a weak gamma-ray source.
In contrast, a high-luminosity source like 3C~279 would be observed either 
during a flare, in which case it would appear quite luminous, or else 
tend to be undetected.
Indeed, many bright LBLs have never been detected with EGRET.
(Note that high-luminosity blazars have a higher average redshift 
than low-luminosity blazars, due to their lower space density, and so 
are less likely to be close enough to detect during quiescent periods.)

The net effect of measuring systematically much-high\-er-than-aver\-age gam\-ma-ray 
lu\-mi\-no\-si\-ties in high-lu\-mi\-no\-si\-ty bla\-zars, and rough\-ly-aver\-age-qui\-e\-scent
gam\-ma-ray lu\-mi\-no\-si\-ties in low-lu\-mi\-no\-si\-ty bla\-zars, is to produce the observed 
trend of $L_C/L_s$ with $L_{bol}$, at least qualitatively.
In the well-known multi-epoch SED of 3C~279 (e.g., \cite{aew1,aew2}), 
this is seen explicitly: during the highest state, the Compton ratio is 
between 10 and 100, while during the lowest state, it is 1 or perhaps 
less (it is poorly determined because the synchrotron peak for this source
is in the far-IR and was seldom observed in the key multiwavelength campaigns).

Thus both biases, the greater beaming for EC compared to SSC, and the
tendency to detect single flares versus multiple flares depending on 
luminosity, lead in the same direction, toward higher Compton ratios in
high-luminosity sources. Whether these biases can explain the observed 
trend quantitatively is not yet clear, but they need to be evaluated in
within the context of any blazar paradigm (necessarily in a model-dependent 
way).

\section{Properties of TeV-Bright Blazars}

\subsection{What is known so far}

What then are the general properties of the TeV-bright blazars?
So far, only two, Mkn~421 and Mkn~501, have been studied in detail, and
few additional sources have been reported
(1ES~2344+514, \cite{cat1}; 
PKS~2155--304, \cite{turv98};
1ES~1959+050, \cite{kajino}).
With only two sources reported more than once in the
literature, we know little about TeV blazars as a class.

The SEDs of the two well-studied TeV-bright blazars have the usual 
shape for blue blazars. For Mkn~421, the synchrotron peak 
is at or slightly above 10$^{17}$~Hz, and the Compton peak (including
the effect of the Klein-Nishina cross section, which suppresses the Compton 
component at lower frequencies than would be the case in the Thomson limit)
occurs just below 1~TeV. The observed Compton ratio is $\simgt1$ 
(\cite{TMG}). 
Mkn~501 is slightly different, with a typically lower Compton ratio, 
although this impression is strongly influenced by the bright flare 
in spring 1997, when the synchrotron peak increased to $\sim100$~keV 
(\cite{pian98,cat97,aha97}).

Note that, as \cite{TMG} shows so clearly,
the TeV emission comes largely from low-energy electrons 
scattering low-energy photons (below the synchrotron peak), 
with high-energy electrons 
(those with $\gamma_e > \gamma_{peak}$) contributing to a hard tail
above the Compton peak. This means the TeV emission reflects closely the
electron spectrum and variability, and is relatively insensitive to
changes in the seed photon flux (which changes slowly at frequencies
below the synchrotron peak; \cite{UMU}). 

\subsection{HEGRA survey}

Following the general paradigm, we expect blazars with high-energy
synchrotron peaks, the HBL/blue blazars, 
to be bright at TeV energies.
Accordingly, we undertook a survey with HEGRA of likely TeV candidates
(with Ron Remillard and Felix Aharonian),
starting from a list of X-ray-peaked, X-ray-bright, nearby BL Lacs.
(A few LBL, typically well-known BL Lacs, were also observed.) Because
these sources are X-ray bright, they are detectable with the RXTE 
All Sky Monitor (ASM), at
least in their higher states. If these objects are like Mkn~421, then
the 2-10~keV X-ray emission is near the peak synchrotron output and the TeV 
emission is near the peak Compton output; thus, the HEGRA measurements
are a probe of the Compton ratio. Our goal was to measure this ratio,
or a limit to it, for blazars generally. 

\begin{figure}
\centerline{\psfig{file=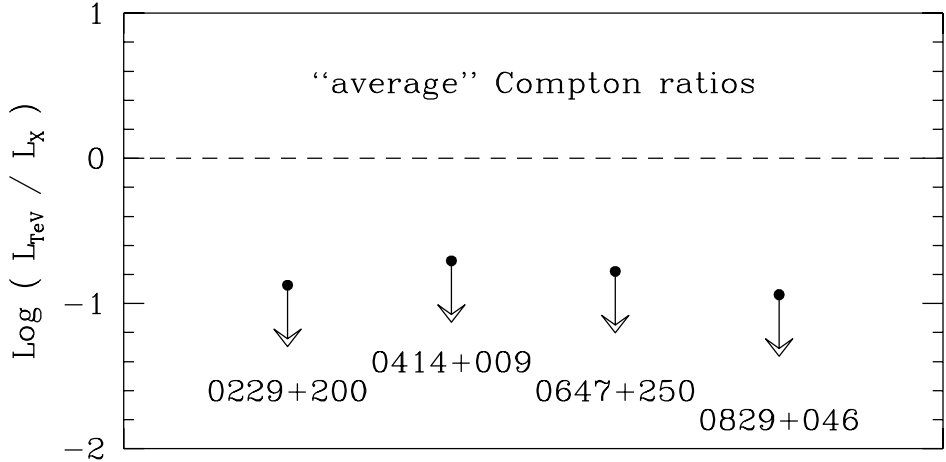,width=0.9\linewidth}}
\caption{Upper limits to the Compton ratio, $L_C/L_s$, of X-ray bright blazars.
The average value is significantly below unity, implying that a relatively 
small fraction of the bolometric luminosity is typically radiated in gamma-rays,
at least in the quiescent state. Estimated from HEGRA upper 
limits (\cite{cecile}) and simultaneous RXTE ASM fluxes, assuming the
Compton and synchrotron peaks lie near 1~TeV and 1~keV, respectively.}
\label{fig:limits}
\end{figure}

In the end, we observed 10 sources, including 7 HBL and 3 LBL. 
None were detected and there are only upper limits for the TeV flux
(\cite{cecile}). 
Because HEGRA sensitivity is $\sim10^{-11}$~ergs~cm$^{-2}$~s$^{-1}$ and
ASM sensitivity is $\sim10^{-10}$~ergs~cm$^{-2}$~s$^{-1}$,
we can in principle, using simultaneous ASM observations, probe an
interesting regime in Compton ratio, as low as $L_C/L_s \sim 1/10$. 
Unfortunately,
six of the sources were also below the ASM detection limit at the epoch
of the HEGRA observations, thus we have only four estimates of the
Compton ratio, from 3 HBL and 1 LBL (see Fig.~\ref{fig:limits}). 
The best estimates of
the ratios all lie well below 1, closer to 0.1 or 0.2. The 95\%
confidence upper limits range from 0.35 down to 0.18. Either the
typical ratios in blazars are below 1, or the spectra of these four
objects differ from Mkn~421. Specifically, if the 2-10 keV ASM band
is near the peak of the synchrotron component but the 1-10 TeV HEGRA
band lies above the peak of the Compton component, then the measured
limit underestimates the Compton ratio.

From this limited survey we infer that the typical Compton ratios of TeV-bright 
blazars may be closer to $L_C/L_s \sim 0.1$ than $\sim1$. 
This implies different source parameters than for Mkn~421, 
for which $L_C/L_s$ is 
usually $\sim1$. For example, \cite{TMG} derived constraints on 
the magnetic field and the Doppler factor of Mkn~421 in the context 
of a simple homogeneous emitting volume. If the value of $L_C/L_s$
is lower by a factor of $\sim10$, the magnetic field and/or
the Doppler factor must be higher by a factor of a few.

\subsection{Directions for future TeV surveys }

Our TeV survey was limited not by the lack of HEGRA detections but by
the lack of simultaneous X-ray detections (the ASM is a factor of a few 
less sensitive than required). In the future, in order
to estimate the Compton ratio in an unbiased way, the most fruitful approach
would be to dedicate pointed RXTE or SAX or ASCA observations to 
a well-chosen HBL sample, observing simultaneously in the TeV. X-ray
detection would be assured, so that any TeV limit would be significant,
and the X-ray spectra would allow determination of the synchrotron peak, 
so the TeV limits could be interpreted with less ambiguity. 

Alternatively, to improve the chances of a TeV detection, one could
make the plausible link between X-ray flaring and TeV flaring, as
observed explicitly in Mkn~421 and Mkn~501, and thus try to 
trigger on X-ray flares. 
From the ASM data (http://space.mit.edu/XTE/ASM\_lc.html)
it is clear that many blazars have relatively short-lived high states.
Thus one would need to trigger with very little delay, perhaps responding
within 1~day to ASM detections of bright states. With the automated state
of data processing in the ASM, such rapid notification to TeV 
observers should be routinely possible.

\subsection{New observations of Mkn~421 in spring 1998}

\begin{figure}
\centerline{\psfig{file=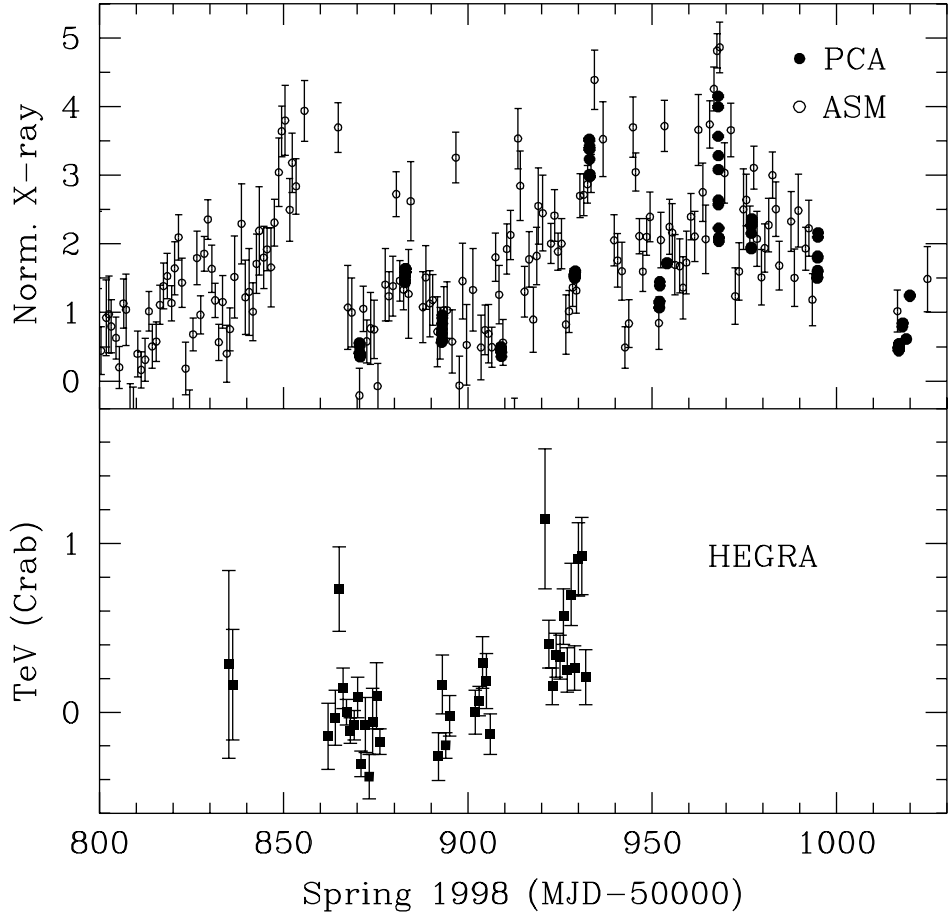,width=0.9\linewidth}}
\caption{X-ray and TeV light curves of Mkn~421 during a flaring state in
spring 1998 (MJD 50800 is 1997 December 28). 
X-ray data are from the RXTE Proportional Counter Array (PCA),
obtained during a TOO monitoring program with Ron Remillard, PI, and the
RXTE All-Sky Monitor (ASM). Simultaneous HEGRA data 
are reproduced here with permission
(HEGRA collaboration, to be published).}
\label{fig:421lc}
\end{figure}

Since Mkn~421 is a very bright source at X-ray and TeV energies, it is
an obvious target for monitoring, 
even recognizing that the results are likely biased toward the flaring 
state. When RXTE ASM data indicated that Mkn~421 was flaring in February 1997,
we triggered a Target of Opportunity (TOO), 
observing roughly every 10 days with the RXTE PCA
through July 1998, with a few additional observations in August 1998. 
Daily less-sensitive X-ray observations are available from the ASM. 
During much of this time HEGRA (and other TeV telescopes) were also 
observing, except during bright time.
Figure \ref{fig:421lc} shows the X-ray and TeV light curves. 

In the X-ray, the PCA (filled circles) detects short time scale variability,
while the light curve of daily ASM averages (open circles) samples 
the longer time scale variability; the two agree well.
In early June 1998 (around MJD 50970), the PCA light curve shows a
steep decline in X-ray flux, by more than a factor of 2 in 5 hours.

The TeV and X-ray fluxes appear well correlated, consistent with
the zero lag (within 1 day) found from more extensive X-ray and TeV
light curves of Mkn~421 (\cite{K98}). With additional TeV data from the 
June 1998 flare, 
it should be possible to comment on lags at shorter time scales.
This is certainly
in accord with the blazar paradigm discussed above, where a single
electron population produces the X-rays via synchrotron radiation and
the TeV emission via Compton scattering of soft photons (probably low-frequency
synchrotron radiation).

\subsection{Multiwavelength observations of the well-studied BL Lac object 
PKS~2155--304}

Only a handful of blazars have been well-monitored simultaneously
in multiple spectral bands, the GeV-bright blazar 3C~279 and a few others,
and the TeV-bright blazars Mkn~421 and PKS~2155--304. While Mkn~421 has also
been monitored at TeV energies (and presumably PKS~2155--304 will be as well
now that southern Cerenkov telescopes are coming on line), PKS~2155--304
has probably the best simultaneous data at optical through X-ray 
energies, which is to say at the high energy end of the synchrotron
component.

There have been four ex\-ten\-sive
mul\-ti\-wave\-length cam\-paigns to ob\-serve PKS 2155--304 with
multiple satellites. In November 1991 (\cite{E95}), 
Rosat (soft X-ray) and IUE (UV and optical) light curves
showed closely correlated variations of modest, wavelength-independent
amplitude ($\simlt 30$\%), with the X-rays leading the UV by a few hours. 
A second campaign in May 1994 using ASCA, Rosat, EUVE, IUE (without the 
optical data), and ground-based telescopes showed altogether different
results: a strong isolated flare in each band,
with much larger amplitude at shorter wavelengths, and with 1-day delays
between X-ray and EUV, and between EUV and UV wavelengths (\cite{U97}). 
While such
disparate behavior might intuitively suggest quite different properties
in the emitting plasma, in fact, \cite{GeorgA}
explain much of the observed multiwavelength behavior starting from the
same underlying jet and varying only the electron injection event.

\begin{figure}
\centerline{\psfig{file=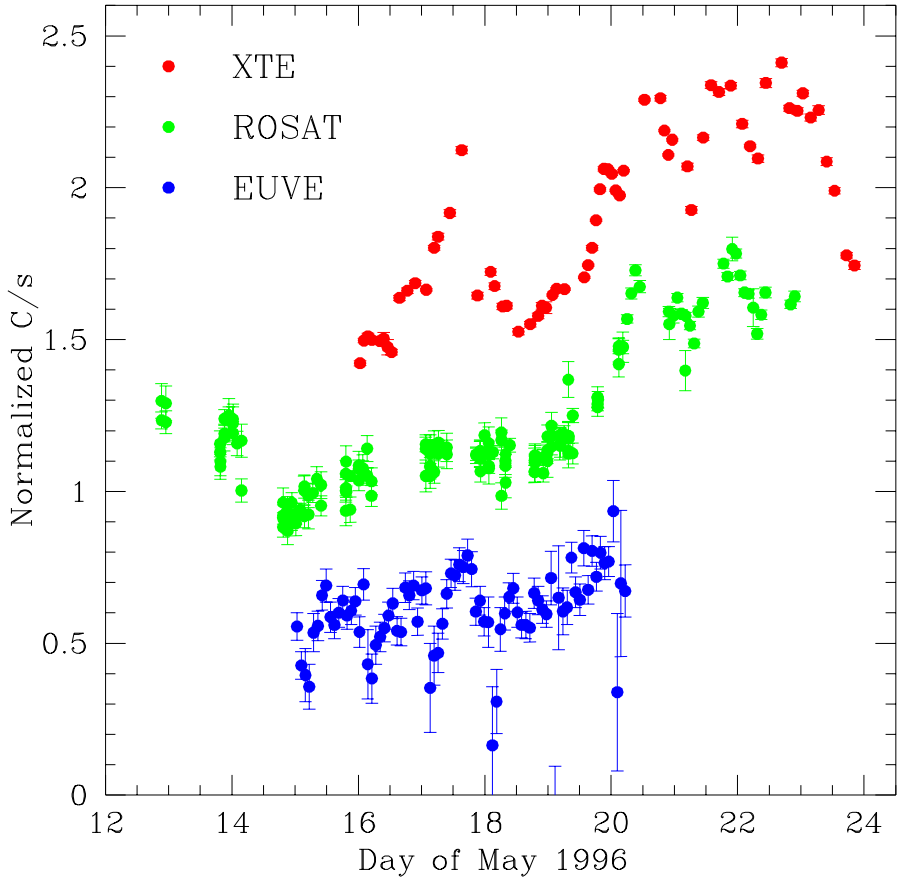,width=0.9\linewidth}}
\caption{Multiwavelength light curves of the TeV blazar PKS~2155--304 in
May 1996 (\cite{S99}). Each band is normalized to its own average, then 
plotted with arbitrary offset, with XTE at the top, ROSAT in the middle,
and EUVE below. As has been seen before, the amplitude increases with 
energy and variations are generally well correlated, with little or no lag.
However, the large flare seen with XTE on May 17, which appears less 
prominently 
in the EUVE light curve, has no counterpart in the ROSAT light curve. 
The light curves of PKS~2155--304 in this third multiwavelength campaign 
differ significantly from those in both the first and second campaigns
(\cite{E95,U97}, obviating simple interpretations of any one
multiwavelength data set.}
\label{fig:2155lc}
\end{figure}

A third multiwavelength monitoring campaign in May 1996, involving
RXTE, Rosat, and EUVE (and lacking IUE because of a spacecraft failure
a few weeks earlier) shows yet another result (\cite{S99,U98}),
shown in Figure~\ref{fig:2155lc}.
The RXTE PCA and Rosat
PSPC light curves are well correlated in the second half of the 
12-day observation, with little or no lag,
but near the beginning, a large flare in the RXTE
data is not seen in the Rosat data (although there is a small gap in
the ROSAT data). A fourth campaign, in November 1996,
with RXTE and Rosat, again shows good correlation as far as the more
limited data allow (\cite{S99,U98}).

These disparate results illustrate the danger of drawing strong
conclusions from single-epoch multiwavelength campaigns, not to mention
the danger of extrapolating from one or two sources to all blazars.

\subsection{The unusual spectral behavior of Mkn~501}

Others have referred to the extraordinary flaring of Mkn~501 in April 1997,
with strong TeV flux and X-ray spectrum hardening in a previously
unobserved way (\cite{pian98,cat97}).
Pian and collaborators observed Mkn~501 again with BeppoSAX
in April 1998, and found that while the X-ray flux remained quite high
and was similar to that seen in April 1997, 
the synchrotron peak was at $\sim10$-20~keV,
still harder than typical but a factor of $\sim10$ lower in energy than
at the peak of the 1997 flare (\cite{pian99}).

Thus the characteristic electron energy is somewhat lower
but there is clearly ongoing acceleration, 
sustained over more than one year, as evidenced by the
continuous RXTE ASM light curves. Apparently the time scales in Mkn~501
are longer than in Mkn~421, at least at the present epoch, despite their
similar luminosities and spectral properties.

\section{What Kind of Jets Does Nature Make?}

\subsection{The extrema of jet physics}

We have described how the spectral energy distributions of blazars
differ markedly between the low-luminosity sources discovered primarily
in X-ray surveys and the high-luminosity sources discovered primarily
in radio surveys. Whether the origin of the SED differences is due
to electron cooling or something else, inevitably the jet
physics in the two types of blazar must differ. In fact, given the
selection biases, there must be a distribution of SEDs, and therefore
a range of typical electron energies and/or jet physics, as indeed has been
found recently in surveys selected to favor intermediate objects
(\cite{Sally,P98}).

Clearly nature makes jets with a range of properties. Perhaps the
process of jet formation varies, or the effect of circumnuclear
environment affects the eventual jet properties, or some combination
of the two (e.g., \cite{Bick95}).
Since jet formation is a fundamental
question, it is of considerable interest to understand the IMF, as it were,
of jets. 

\subsection{Number densities of red and blue blazars}

In fact, the relative number of red blazars and blue blazars is
very poorly known (\cite{UP95}).
X-ray surveys, which find
the blue blazars because of their high-frequency synchrotron peaks,
span a large range in flux but a relatively small range in luminosity.
Radio surveys, which pick out red blazars because of their low-frequency
peaks, span a small range in flux but a rather larger range in luminosity.
Both are undoubtedly biased, but to find the absolute numbers of blue or
red blazars requires correcting either type of survey for the kind of
objects not found in that survey. 

\begin{figure}
\centerline{\psfig{file=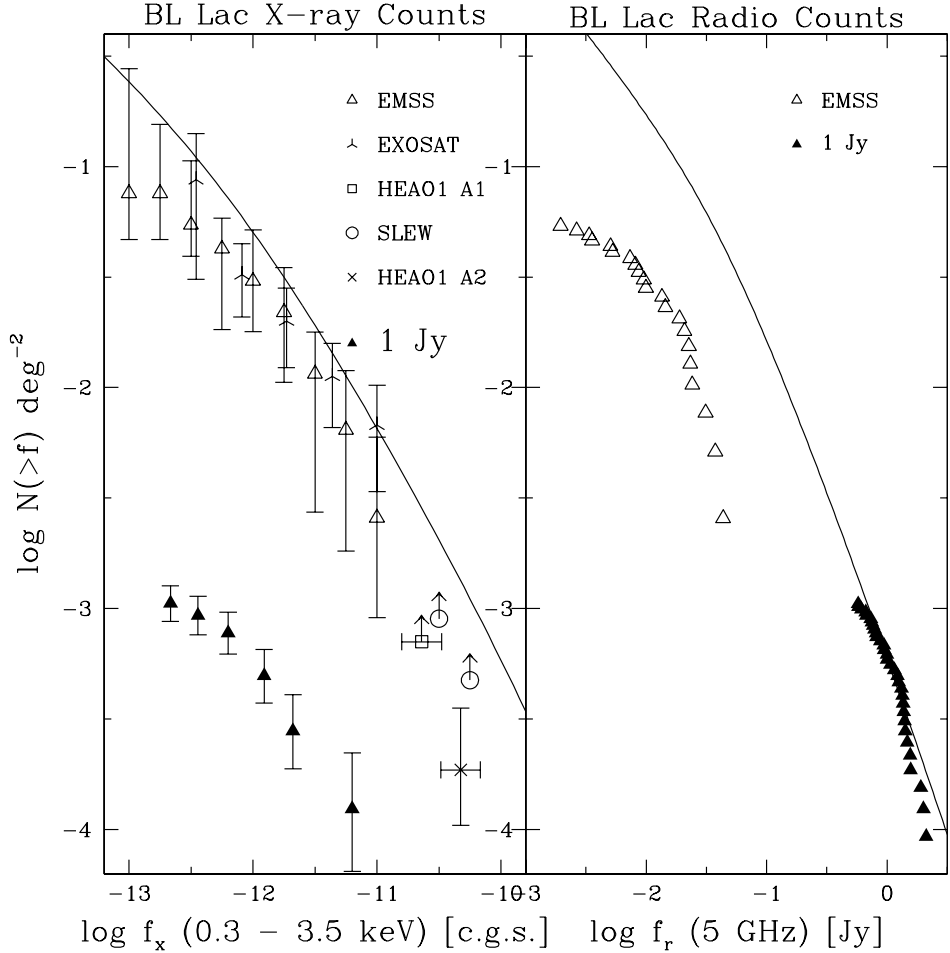,width=0.9\linewidth}}
\caption{The X-ray and radio number counts for blue (open symbols) and
red (filled symbols) BL Lac objects illustrate the ambiguity over which
population is more numerous.
{\it Left ---}
Observed X-ray counts for X-ray-selected samples
are $\sim10$ times higher
than the ``bivariate'' X-ray counts for radio-selected samples
(radio-derived surface density combined with observed X-ray fluxes;
\cite{UPS91}).
{\it Right ---}
Beaming model extrapolation (solid line) of
observed radio counts for radio-selected samples (\cite{UP95})
is $\sim10$ times higher than the ``bivariate'' radio counts
for X-ray-selected samples (X-ray-derived surface density
combined with the observed radio fluxes).
\label{fig:counts}
}
\end{figure}

Deriving the relative density of types of bla\-zars
is in\-ev\-i\-tab\-ly a mo\-del-de\-pen\-dent pro\-cess. 
Figure~\ref{fig:counts} illustrates the dilemma with two extreme cases.
If one assumes that X-ray surveys are unbiased, then the numbers of
X-ray sources is reflected in the source counts, here taken from a 
number of surveys at different flux limits (left panel). To compare the density
of red blazars, we convert the radio counts to X-ray counts via
the average value of $\alpha_{rx}$ appropriate to this sample 
(\cite{UPS91}), and plot these also in the left panel of
Figure~\ref{fig:counts}. The values of $\alpha_{rx}$ are quite steep 
(the X-ray fluxes are low) so the radio counts lie
on the low-flux left side of the plot.
This representation of the source counts suggests 
red blazars are 10 times less numerous than blue blazars.
This is certainly the case for a given X-ray flux but would be true in
a bolometric sense only if the SEDs of red and blue blazars were the
same, which we know very well they are not.

Alternatively, one can assume the radio survey is unbiased, and compare
surface densities of red and blue blazars starting from the 1~Jy
and S4 surveys. In this case, the X-ray samples are converted to radio
fluxes using the average $\alpha_{rx}$ appropriate to this sample
(quite flat, so leading to low radio fluxes). Here the flux ranges
do not overlap at all, but using the extrapolation implied by the 
best-fit beaming model (\cite{PU90,UPS91}; solid line in both
panels), we would conclude that the radio-selected
red blazars are 10 times less numerous than the X-ray-selected blue
blazars.
The truth probably lies between these two extreme views, but at present
it is not constrained.

This has of course been explored in much greater detail, starting from
luminosity functions that agree with observation, making some assumptions,
and calculating absolute numbers (\cite{PG95,Foss}).
We can say that the simple picture
wherein blue blazars are seen at larger angles to the line of sight than
red blazars, on average (\cite{M86}) does not explain the full
range of SEDs (cf. \cite{GeorgB}), and thus the 
implication that blue blazars are more numerous (due to their larger
solid angle) does not seem to be correct.
On the other hand, the assumption that the radio is unbiased 
(\cite{PG95})
and that there is a distribution of spectral cutoffs does explain the
observed X-ray and radio counts, but the calculation is entirely symmetric
with respect to wavelength. One can equally well assume that the X-ray
is unbiased, and that there is a range of spectral cutoffs toward longer
wavelengths, and again the X-ray and radio counts can be matched.

This ambiguity results because present samples are small and, particularly
in the radio, have relatively high flux limits. Even slightly more sensitive 
radio surveys should be able to verify or disprove these two contradictory
hypotheses.

\section{Summary}

A viable blazar paradigm can be constructed from the observed multiwavelength
properties of blazars. All blazars would have synchrotron and 
inverse-Compton-scattered components, 
with external seed photons dominating
the gamma-ray component in high-luminosity sources (the EC model)
and synchrotron seed photons dominating in low-luminosity sources
(the SSC model). 
In both cases, the X-rays are likely to be dominated by Compton-scattered 
synchrotron (SSC) photons.

Empirical relations between source parameters and luminosity suggest electron
acceleration may be universally the same independent of luminosity,
whereas cooling is strongly luminosity-dependent. The more luminous
sources, with their high ambient photon densities, have much greater
cooling and therefore lower average electron energies, and therefore
lower frequency synchrotron and Compton peaks, as well as higher
ratios of Compton to synchrotron radiation. Several selection effects
could exaggerate this trend, however.

The multiwavelength properties of TeV-bright blazars are beginning
to be well studied. However, the average properties of blue blazars
may be different from those of the well-studied TeV blazars Mkn~421
and Mkn~501. For example, a small survey with HEGRA suggests the 
typical Compton ratio may be closer to $L_C/L_s \sim 0.1$ than 1. Further
TeV observations, combined with more sensitive simultaneous
X-ray observations, are needed to confirm this suggestion. In addition,
X-ray monitoring can be used to pick out flaring blazars as potentially
strong TeV sources; however the trigger times needed are quite short,
less than 1 day. Certainly
in Mkn~421 and Mkn~501, the TeV and X-ray light curves are well correlated, 
with delays $\simlt$~day.

PKS~2155--304, perhaps the best observed blazar at UV-X wavelengths,
illustrates the complexity of blazar variability and the danger of
strong conclusions from limited data sets.
Mkn~501 continues to be in a high state but $\gamma_{peak}$ has decreased,
judging from synchrotron peak frequency, which is a factor of $\sim10$ 
lower than one year earlier.

Finally, jet demographics are an interesting and not yet well 
understood issue. Whether nature forms more high-luminosity jets
than low-luminosity jets, as implied if red blazars are more numerous,
or the opposite, is unclear and is of fundamental physical significance.
This problem can be approached with deeper samples of flat-spectrum
radio sources.

\noindent
Acknowledgements --- It is a pleasure to thank my blazar colleagues,
particularly those with whom I have worked most closely in the past year,
Riccardo Scarpa, Rita Sambruna, Ron Remillard, Elena Pian, 
Laura Maraschi, Gabriele Ghisellini, Giovanni Fossati, 
Annalisa Celotti, and Felix Aharonian,
and to acknowledge the warm hospitality of 
the Center for Astrophysical Sciences at the Johns Hopkins University 
and the Brera Observatory of Milan during my sabbatical. 
We thank the HEGRA collaboration for permission to use the Mkn~421 data 
in advance of publication.
This work was supported in part by NASA grants NAG5-3313 and NAG5-3138.

\end{document}